\documentclass[english]{article}
\usepackage[T1]{fontenc}
\usepackage[latin9]{inputenc}
\usepackage{geometry}
\usepackage{amsmath}
\usepackage{graphicx}
\PassOptionsToPackage{normalem}{ulem}
\usepackage{ulem}

\makeatletter

\newcommand{\lyxaddress}[1]{
	\par {\raggedright #1
	\vspace{1.4em}
	\noindent\par}
}

\makeatother

\usepackage{babel}
\begin{document}
\title{Notes on Hydraulic Fracture Mechanics \thanks{Developed for \textbf{CISM} (Centre International des Sciences Mécaniques)
\textbf{advanced school} on \textbf{\emph{Coupled Processes in Fracture
Propagation in Geo-Materials: From Hydraulic Fractures to Earthquakes}},
Udine Italy, \textbf{June 10-14, 2019}}}
\author{Dmitry I. Garagash}
\maketitle

\lyxaddress{Dalhousie University, Department of Civil and Resource Engineering,
Halifax, Canada}
\begin{abstract}
These notes address mechanics of propagating hydraulic fractures (HF).
In the first part, we focus on how different physical mechanisms (dissipation
in fluid and solid, fluid storage in fracture and its exchange with
permeable rock, crack elasticity) manifest near the fracture tip and
lead to a `zoo' of fracture propagation regimes, as correspond to
different coupling of predominant mechanisms depending on material
parameters and fracture propagation speed. In the second part, we
illustrate how different near-tip regimes dictate the propagation
of finite fractures of simple geometries (e.g. 2D plane-strain, or
3D radial cracks) driven by fluid injection at the center of the crack.
These notes \emph{are }\emph{\uline{not}}\emph{ a review} of the
research on the topic, which has seen tremendous renewed interest
in the last 20 years or so, (for a review see, e.g., \cite{Detournay16});
but rather an attempt to introduce the mechanics and physics of HF
in a simple and hopefully-logical way, from the fracture tip to a
finite fracture propagation. 
\end{abstract}

\section{Boundary Layer Structure Near Tip of Propagating Hydraulic Fracture
via Williams-like expansions}

Here we will explore how the presence of fluid at the propagating
fracture tip may impact the near tip fracture behavior in the context
of limiting propagation regimes dominated by a subset of the physical
processes. These processes can be simply identified with two energy
dissipation mechanisms in the HF:
\begin{itemize}
\item dissipation in generation of new fracture surface, i.e. the classical
fracture energy or toughness, and
\item the viscous dissipation in the fluid flow within the fracture
\end{itemize}
and with two fluid storage mechanisms
\begin{itemize}
\item in the fracture, and
\item in the surrounding permeable rock (leak-off).
\end{itemize}

\subsection{William's solution of elasticity}

Recall Williams' power-law solution for a semi-infinite Mode I (symmetric)
plane-strain crack in infinite elastic domain 
\begin{figure}[tbh]
\includegraphics[scale=0.1]{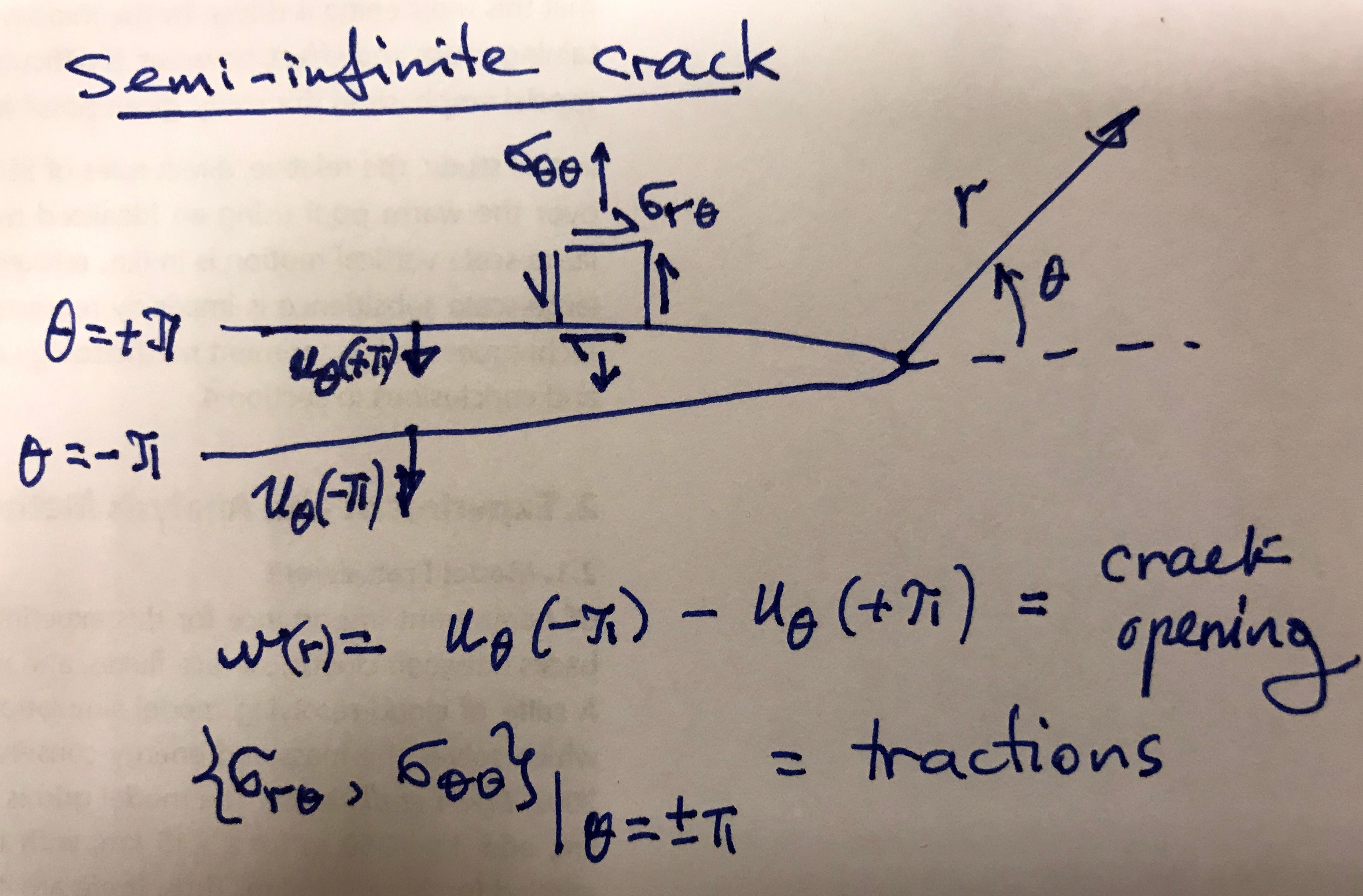}

\caption{Semi-infinite crack and related definitions.\label{fig:semi-inf}}
\end{figure}
 for the Airy Stress Function \cite{williams1952Expansion,Barber10}
\[
\phi=r^{\lambda+1}(A_{1}\cos(\lambda+1)\theta+A_{2}\sin(\lambda-1)\theta)
\]
(a power law form which satisfies the biharmonic equation $\nabla^{4}\phi=0$),
stress
\begin{align}
\sigma_{r\theta} & =r^{\lambda-1}\left(A_{1}\lambda(\lambda+1)\sin(\lambda+1)\theta+A_{2}\lambda(\lambda-1)\sin(\lambda-1)\theta\right)\label{williams_stress}\\
\sigma_{\theta\theta} & =r^{\lambda-1}\left(A_{1}\lambda(\lambda+1)\cos(\lambda+1)\theta+A_{2}\lambda(\lambda-1)\cos(\lambda-1)\theta\right)\nonumber 
\end{align}
and displacements
\begin{align}
u_{r} & =\frac{1}{2\mu}r^{\lambda}\left(-A_{1}(\lambda+1)\cos(\lambda+1)\theta+A_{2}(\kappa-\lambda)\cos(\lambda-1)\theta\right)\label{williams_displ}\\
u_{\theta} & =\frac{1}{2\mu}r^{\lambda}\left(A_{1}(\lambda+1)\sin(\lambda+1)\theta+A_{2}(\kappa+\lambda)\sin(\lambda-1)\theta\right)\nonumber 
\end{align}
where $\kappa=3-4\nu$ (plane strain) and $\mu$ is the shear modulus.
(In plane strain problems, we will make use of the so-called plane-strain
elastic modulus, $E'=2\mu/(1-\nu)$).

This solution can be used to construct a variety of hydraulic fracture
tip solutions characterized by power-law tractions along the fracture
faces. Coefficients $A_{1}$ and $A_{2}$ are recovered as part of
the solution of a particular problem in the following.

\subsection{Fluid flow in a steadily propagating fracture in impermeable rock}

\subsubsection{Impermeable solid}

Consider hydraulic fracture tip propagating at a steady velocity $V$
on assumption that the fluid flow in the fracture is able to ``keep
up'' with the fracture tip. Conservation of mass (volume for an incompressible
fluid) suggests that fluid flow velocity is 
\begin{equation}
v_{f}=V\label{balance}
\end{equation}
where the fluid velocity is given by the Poiseuille's law for pressure-gradient
driven flow 
\begin{equation}
v_{f}=\frac{w^{2}}{\eta'}\frac{dp_{f}}{dr}\label{lub}
\end{equation}
in a channel of aperture 
\[
w=u_{\theta}(-\pi)-u_{\theta}(+\pi),
\]
(Here $\eta'=12\eta_{f}$ is a fluid viscosity parameter and $p_{f}$
is the fluid pressure.)

\subsubsection{Permeable solid (fluid leak-off)}

A fracture pressurized by a fluid at $p_{f}\sim\sigma_{o}$ would
necessarily leak fluid off to the surrounding rock, if a) the rock
is sufficiently permeable and b) the ambient pore fluid pressure in
the rock $p_{o}$ is less than the fracturing fluid pressure, i.e.
$p_{o}<p_{f}$. The latter condition is typically satisfied since
the ambient pore pressure is (much) smaller than the ambient stress,
$p_{o}<\sigma_{o}$.

On assumption $p_{o}\ll\sigma_{o}\sim p_{f}$, the leak-off can be
modeled as from approximately constant fluid pressure step (from $p_{o}$
to $p_{f}\sim\sigma_{o}$) from the moment of the arrival $t\ge t_{o}(x)$
of the fracture tip at a given position $x$ along the fracture path.
Limiting the diffusion considerations to the 1D (the leak-off boundary
layer around the fracture is small compared to a pertinent fracture
lengthscale, e.g. length for finite fractures), the rate of leak-off
is then $g(x)=C'/\sqrt{t-t_{o}(x)}$ where $C'$ is so-called Carter's
leak-off coefficient incapsulating the rock parameters and proportional
to the imposed fluid pressure step at the fracture wall $\approx\sigma_{o}-p_{o}$).

The fracturing fluid balance equation 
\[
\frac{\partial w}{\partial t}+\frac{\partial wv_{f}}{\partial x}+g=0
\]
when rewritten in the frame moving with crack tip, $r=Vt-x$ and $\theta=\pm\pi$,
becomes
\[
V\frac{dw}{dr}-\frac{dwv_{f}}{dr}+g=0
\]
where the leak-off rate can be now expressed as $g=C'\sqrt{V}/\sqrt{r}$
(since the crack arrival time at fixed position $x=Vt-r$ can be expressed
as $t_{o}(x)=r/V$). Integrating the above equation from the tip $r=0$
to some $r>0$ and using tip closure condition $w_{|r=0}=wv_{|r=0}=0$,
we finally recover
\begin{equation}
wv_{f}=wV+2C'\sqrt{Vr}\label{balance_leak}
\end{equation}

Compare this to the impermeable case fluid balance, $v_{f}=V$, to
appreciate that the fluid inflow into the tip region bounded by some
$r$ (i.e. $wv_{f}$) is partitioned between the fluid stored in that
crack tip region (i.e. $wV$) and the fluid leaked-off into the rock
from that region (i.e. $\int_{0}^{r}gdr=2C'\sqrt{Vr}$). Thus, the
right hand side of this equation simply presents the fluid partition
between crack storage and ``rock storage'' (or leak-off).

\begin{figure}[tbh]
(a)\includegraphics[scale=0.095]{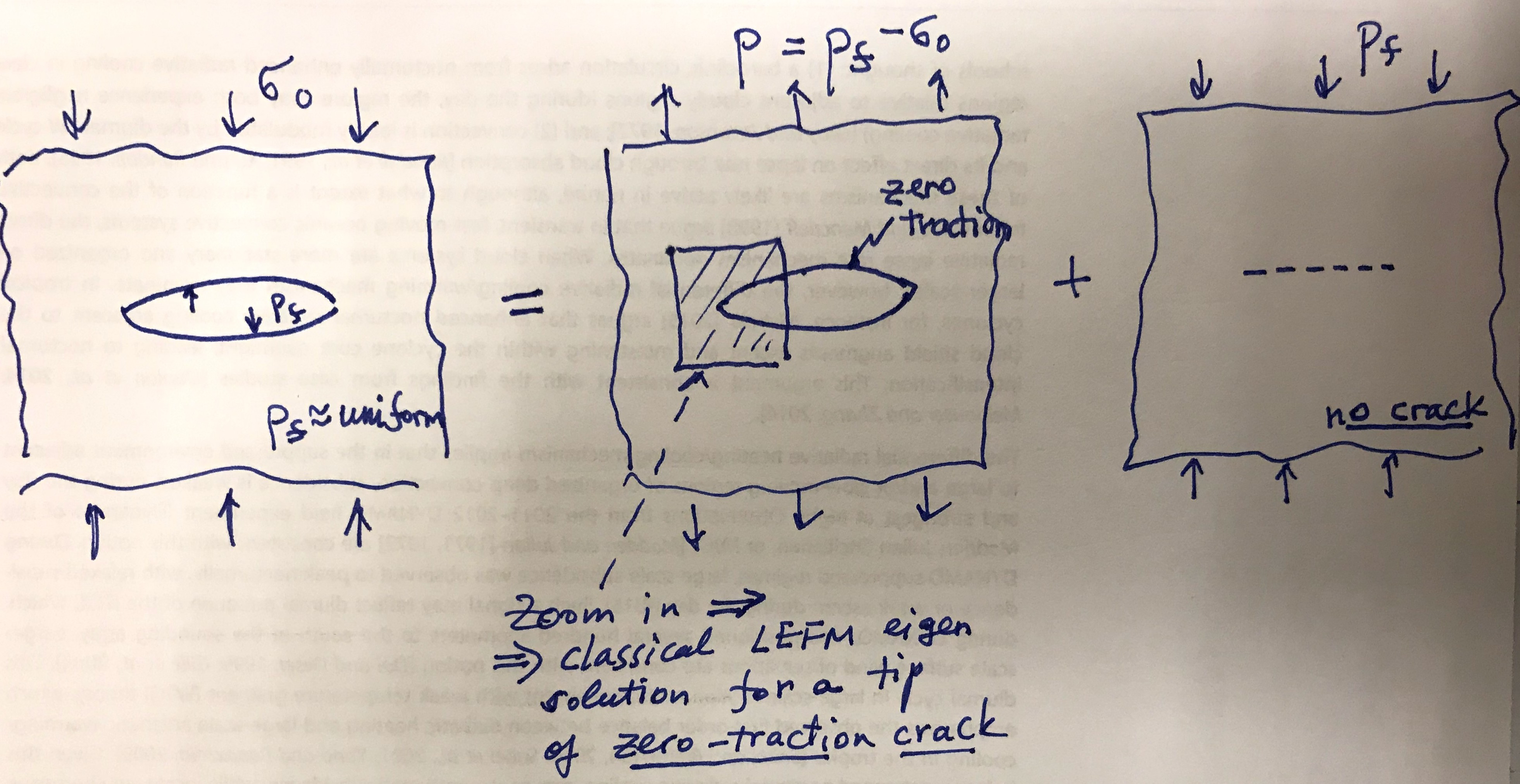}

(b)\includegraphics[scale=0.1]{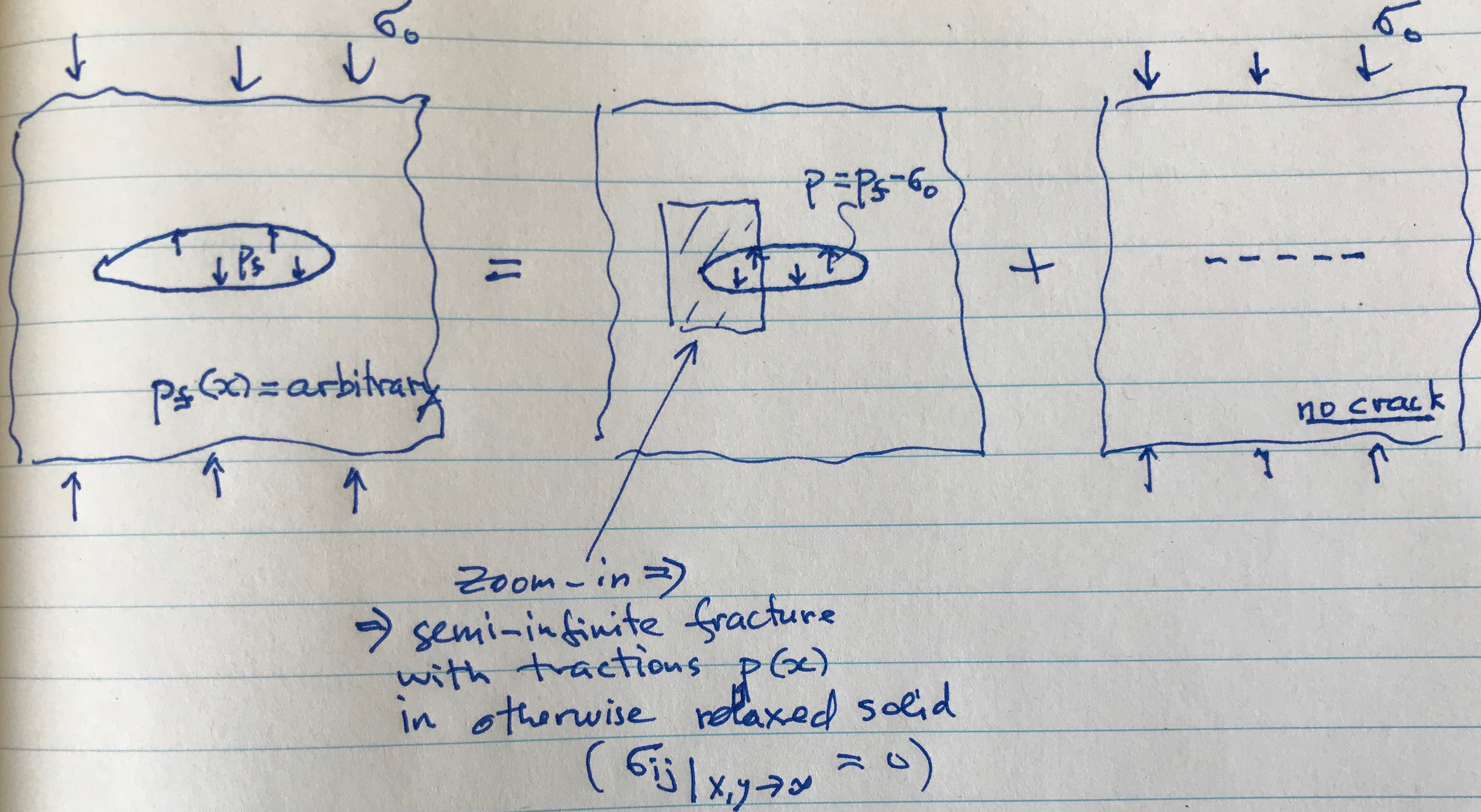}

\caption{Linear decomposition of the HF problem when distribution of normal
traction on the crack walls (fluid pressure) is (a) uniform and (b)
non-uniform.\label{fig:decomp}}
\end{figure}

\subsection{Uniformly loaded HF - Toughness-dominated solution\protect \\
(negligible viscous pressure drop in the crack)}

In the limiting case when the fluid pressure drop in the flow towards
the fracture tip is negligible (e.g., when fluid viscosity $\eta'$
is negligible), i.e. $p_{f}$ is approximately uniform along the crack,
the problem can be decomposed into that of (i) a zero-traction crack
in the elastic domain loaded by tensile far field stress which value
is given by the fluid net pressure, $\sigma_{yy}(x,y\rightarrow\infty)=p=p_{f}-\sigma_{o}$;
and (ii) the uniformly stressed elastic domain (without the crack),
$\sigma_{yy}(x,y)=-p_{f}$ (see Figure \ref{fig:decomp}a).

Zooming into the crack tip region of the crack in sub-problem (i),
we recover a semi-infinite, traction-free crack (Figure \ref{fig:semi-inf})
with
\[
\sigma_{\theta\theta}(\theta=\pm\pi)=0\qquad\sigma_{r\theta}(\theta=\pm\pi)=0
\]

This is of course a familiar (by now) problem of the zero-traction
crack tip field, which solution follows from evaluating the above
boundary conditions using the Williams' solution (\ref{williams_stress}).
This is an eigenvalue problem, which solution leads to the LEFM power-law
$\lambda=1/2$ and a relation between the two unknown coefficients
$A_{1}$ and $A_{2}$ in (\ref{williams_stress}). Eliminating $A_{2}$
and redefining the unknown $A_{1}\Rightarrow K_{I}/(3\sqrt{2\pi})$,
we get the crack tip solution for the zero-traction semi-infinite
crack in the form
\[
\left(\begin{array}{c}
\sigma_{r\theta}\\
\sigma_{\theta\theta}
\end{array}\right)=\frac{K_{I}}{\sqrt{2\pi r}}\left(\begin{array}{c}
\sin\theta\,\cos^{2}\frac{\theta}{2}\\
\cos^{3}\frac{\theta}{2}
\end{array}\right)\qquad\left(\begin{array}{c}
u_{r}\\
u_{\theta}
\end{array}\right)=\frac{4K_{I}}{E'(1-\nu)}\sqrt{\frac{r}{2\pi}}\left(\begin{array}{c}
\cos\frac{\theta}{2}\,\left(1-\frac{1+\cos\theta}{4(1-\nu)}\right)\\
-\sin\frac{\theta}{2}\,\left(1-\frac{1+\cos\theta}{4(1-\nu)}\right)
\end{array}\right)
\]
$K_{I}$ with units {[}Pa$\sqrt{\text{m}}${]} is the so-called mode
I stress intensity factor (SIF), as it gives the ``strength'' of
the $1/\sqrt{r}$ stress singularity of the stress ahead of the tip.
More specifically, the tensile singular stress ahead of the fracture
tip, acting to ``open''/advance the crack along its plane ($\theta=0$)
is varying with distance $r$ ahead of the tip as
\[
\sigma_{yy}(r,\theta=0)=\frac{K_{I}}{\sqrt{2\pi r}}
\]
while the crack opening is varying with distance $r$ behind the crack
tip as 
\begin{equation}
w(r)=u_{\theta}(r,-\pi)-u_{\theta}(r,+\pi)=\frac{8K_{I}}{E'}\sqrt{\frac{r}{2\pi}}\label{k-1}
\end{equation}

When zooming to the traction-free fracture tip in the subproblem (i)
of Figure \ref{fig:decomp}a, we essentially ``lost'' the information
about the global (far-field) tensile loading, which keeps the finite
fracture open/advancing, and, thus, determines the magnitude of the
stress intensity factor. In other words $K_{I}$ is inherently global
loading characteristic, which provides a link between the finite fracture
and its (far field in our case) loading and the its near tip behavior.

We will be able to evaluate $K_{I}$ for a given global fracture problem
of interest, once the elasticity equation for a crack of a finite
geometry is formulated. Notwithstanding, for \emph{propagating} fractures,
the value of $K_{I}$ is constrained by the propagation criterion,
which matches the potential elastic energy ($d\varPi$) that can be
released into the crack tip region upon advancing the crack a unit
distance (a unit surface area, $d\Sigma$) to the fracture energy
$G_{c}$, a material property corresponding to the energy dissipation
involved in creating a unit of (two) new surfaces. The former, referred
to as the energy release rate $G$, is actually uniquely defined by
the elastic crack field \cite{Rice68}
\begin{equation}
G=-\frac{d\varPi}{d\Sigma}=\lim_{\Delta\ell\rightarrow0}\frac{1}{2\Delta\ell}\int_{0}^{\Delta\ell}\sigma_{yy}^{(0)}(r,\theta=0)\,w^{(\Delta\ell)}(\Delta\ell-r)\,dr\label{ERR}
\end{equation}
where the ``$(0)$'' and ``$(\Delta\ell)$'' refer to the fields
before and after crack tip advance by $\Delta\ell$. Evaluating, leads
to a simple relation of the energy release rate to the stress intensity
factor at the crack tip
\[
G=\frac{K_{I}^{2}}{E'}
\]
(The result, which up to a numerical prefactor, follows from units
considerations, where the energy release rate G {[}J/m$^{2}$=Pa m{]}
is ought to be expressed in terms the only two dimensional parameters
defining the LEFM crack tip fields, i.e. $K_{I}$ {[}Pa$\sqrt{\text{m }}${]}
and elastic modulus $E'$ {[}Pa{]}).

The propagation criterion $G=G_{c}$, can be reformulated in terms
of the critical value of the stress intensity factor (referred to
as material ``toughness'') $K_{Ic}=\sqrt{G_{c}E'}$, as 
\begin{equation}
K_{I}=K_{Ic}\label{propa}
\end{equation}
and corresponding tip stress/displacement fields uniquely defined
for propagating fracture.

Defining the lenghthscale 
\[
\ell_{k}=\left(\frac{K'}{E'}\right)^{2}
\]
where for brevity
\[
K'=\sqrt{\frac{32}{\pi}}K_{Ic}
\]
the opening distribution (\ref{k-1}) with (\ref{propa}) for the
toughness-dominated propagating fracture can be rewritten as
\begin{equation}
w=\ell_{k}^{1/2}r^{1/2}\label{k}
\end{equation}

\subsection{Non-uniformly loaded HF (negligible fracture toughness)}

\subsubsection{Impermeable solid}

Consider now the case where viscous dissipation in the fluid flow
is important, and suggest that the net fluid pressure $p(r)=p_{f}(r)-\sigma_{o}$
in the subproblem (i) of Figure\ref{fig:decomp}b (i.e. hydraulic
fracture driven by $p$ in otherwise relaxed solid) is some power-law
of the distance $r$ from the tip, in the crack tip vicinity (modeled
as a semi-infinite crack), 
\begin{equation}
p(r)=-B\,r^{\lambda-1}\label{pm}
\end{equation}
where prefactor $B$ and power law exponent $\lambda$ are unknowns.

Corresponding traction boundary conditions on the semi-infinite crack
walls are 
\[
\theta=\pm\pi:\qquad\sigma_{r\theta}=0\qquad\sigma_{\theta\theta}=-p(r)=B\,r^{\lambda-1}
\]
Corresponding Williams' solution coefficients satisfying the above
boundary conditions are
\[
A_{1}=-B\frac{\lambda-1}{\lambda+1}\frac{1}{2\lambda\cos\pi\lambda}\qquad A_{2}=-B\frac{1}{2\lambda\cos\pi\lambda}
\]
and the crack opening
\begin{equation}
w(r)=u_{\theta}(r,-\pi)-u_{\theta}(r,+\pi)=\frac{B}{E'}\varphi(\lambda)\,r^{\lambda}\label{wm}
\end{equation}
where $\varphi(\lambda)=4\lambda^{-1}\tan(1-\lambda)\pi$.

Plugging the above power laws for the net pressure and opening into
the fluid lubrication equation for \emph{impermeable} solid, $v_{f}=V$,
with $v_{f}=(w^{2}/\eta')(dp_{f}/dr)$ and $dp_{f}=dp$ (uniform far
field confining stress $d\sigma_{o}/dr=0$), we recover the power
law $\lambda=2/3$ and the prefactor $B=\delta_{m}(V\eta'E'^{2})^{1/3}$
with $\delta_{m}=6^{-2/3}$.

Defining lengthscale
\[
\ell_{m}=\frac{V\eta'}{E'}
\]
final expressions for the near HF tip viscous-dominated solution \cite{DeDe94}
\begin{equation}
\sigma_{yy}(\theta=\pm\pi)=-p=\delta_{m}\,E'\,(\ell_{m}/r)^{1/3}\qquad w=\delta_{m}\varphi(2/3)\,\ell_{m}^{1/3}r^{2/3}\label{m}
\end{equation}

This solution suggests that the viscosity-dominated propagating crack
tip is sharper/narrower than the toughness dominated one ($r^{1/2}\gg r^{2/3}$
when $r\rightarrow0$) and that the fluid undergoes \emph{suction}
($p$ and $p_{f}$ are negative close to the tip) diverging towards
the fracture tip.

For the stress ahead of the fracture tip we can recover from the solution
$\sigma_{yy}(\theta=0)=2B/r^{1/3}$, which suggests a tensile singularity
at the tip approached from the intact solid, but a weaker one than
the LEFM one ($\sim1/r^{1/2}$). This state of affairs naturally implies
that the energy release rate (\ref{ERR}) at the viscosity-dominated
crack tip is zero, $G=0$, and that in view of the fracture propagation
criterion $G=G_{c}$, the explored solution is strictly valid only
in the zero-toughness limit. As we will observe in the following,
this seeming contradiction can be reconciled, and the two limiting
(toughness- and viscosity-dominated) solution arise as the near (small
$r$) and far (large $r$) fields of the more general HF tip solution
to be discussed in the following.

\subsubsection{Permeable solid (leak-off)}

Consider now the case of non-negligible fluid leak-off into the permeable
solid around it. To appreciate the importance of the leak-off to the
fracture propagation, assume for a moment that the leak-off effect
is small and use the zero-leak-off viscosity-dominated solution derived
in the above to evaluate the terms in the right hand side of the fluid
balance equation (\ref{balance_leak}), i.e. the partition between
the crack fluid storage $wV$ and the leak-off $\int_{0}^{r}gdr=2C'\sqrt{Vr}$.
The former is $\propto r^{2/3}$ while the latter $\propto r^{1/2}$,
suggesting that the fluid balance in the crack will be dominated by
the crack-storage at large distances from the tip while the leak-off
(i.e. fluid storage in the rock) will dominate at small distances
from the tip.

The near-field, leak-off dominated solution can be recovered by the
same solution procedure as in the impermeable case above, when applied
to the leak-off dominated fluid balance equation end-member $wv_{f}=2C'\sqrt{Vr}$
(i.e. when neglecting the storage term $wV$ in the right hand side)
with $v_{f}=(w^{2}/\eta')(dp/dr)$. Plugging the power law forms for
the net pressure (\ref{pm}) and crack opening (\ref{wm}) into the
fluid balance end-member, we recover the power law $\lambda=5/8$
and the prefactor $B=\delta_{\widetilde{m}}\,\left(C'\sqrt{V}\eta'E'^{3}\right)^{1/4}$
with $\delta_{\widetilde{m}}=\left(\frac{125}{6144}\tan^{3}\frac{\pi}{8}\right)^{1/4}$.

Defining lengthscale
\[
\ell_{\widetilde{m}}=\left(\frac{C'\sqrt{V}\eta'}{E'}\right)^{2/3}
\]
we can write the final expressions for the near HF tip viscous-dominated
solution under conditions of dominant leak-off \cite{Leno95}
\begin{equation}
\sigma_{yy}(\theta=\pm\pi)=-p=\delta_{\widetilde{m}}\,E'\,(\ell_{\widetilde{m}}/r)^{3/8}\qquad w=\delta_{\widetilde{m}}\varphi(5/8)\,\ell_{\widetilde{m}}^{3/8}r^{5/8}\label{m_tilde}
\end{equation}

Similarly to the solution for the viscosity-dominated crack tip in
the impermeable solid, the leak-off (permeable solid) dominated solution
corresponds to the zero energy release rate, and thus strictly realized
in the zero-toughness

\subsection{General HF tip solution \cite{Garagash09,Garagash11}}

In the above, we have recovered three limiting solutions for the problem
of semi-infinite hydraulic fracture propagation, recounted below for
their fracture opening distributions:
\begin{itemize}
\item the toughness-dominated ($k$) solution ($\eta'=0$), $w=\ell_{k}^{1/2}r^{1/2}$
\item the viscosity-storage-dominated ($m$) solution ($K'=0$, $C'=0$),
$w\sim\ell_{m}^{1/3}r^{2/3}$
\item the viscosity-leak-off-dominated ($\widetilde{m}$) solution ($K'=0$,
$C'\rightarrow\infty$), $w\sim\ell_{\widetilde{m}}^{3/8}r^{5/8}$
\end{itemize}
with their respective lengthscales $\ell_{k}$, $\ell_{m}$, and $\ell_{\widetilde{m}}$.

\begin{figure}[tbh]
\includegraphics[scale=0.45]{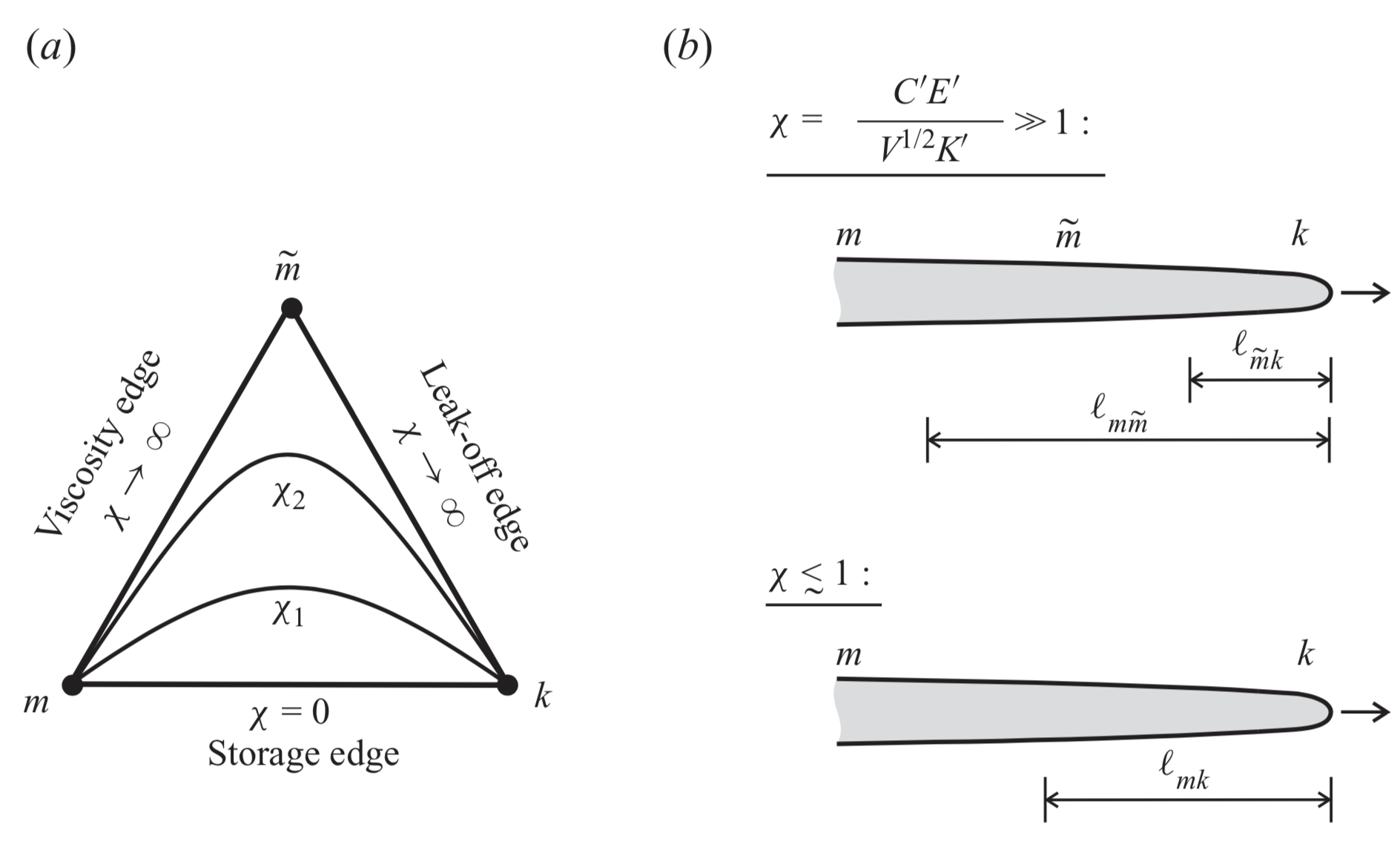}

\caption{Semi-infinite HF parametric space (a), and the structure of the small
($\chi\ll1$) and large $(\chi\gg1$) leak-off solutions, after \cite{Garagash11}.\label{fig:space}}
\end{figure}

The structure of the general solution (i.e. for arbitrary set of values
of fluid viscosity $\eta'$, solid toughness$K_{Ic}$, and fluid leak-off
$C'$ parameters), is hinted by the order of the corresponding power
law exponents in the limiting expressions for the crack opening
\[
1/2<5/8<2/3
\]
which suggests that if all three (or any two of the three) limiting
regimes are to be realized in the general hydraulic fracture solution,
they would appear/dominate at different spatial scales (distances
from the tip): the $k$-solution in the near field ($r\rightarrow0$),
the $m$-solution in the far field ($r\rightarrow\infty$), and the
$\widetilde{m}$-solution possibly in the intermediate field (i.e.
at intermediate distances from the tip)
\begin{equation}
k\rightarrow(\widetilde{m})\rightarrow m\label{solution}
\end{equation}
( arrows indicate the transition with distance away from the fracture
tip). This solution structure can be visualized in a triangular parametric
diagram $m\widetilde{m}k$ \cite{Garagash11}, where the vertices
correspond to the three limiting solutions, and a trajectory reflect
a progression of the solution with the increasing distance from the
tip, always starting from $k$-vertex and ending in the $m$-vertex
(Figure \ref{fig:space}).

The edges of this parametric space correspond to the propagation regimes
dominated by \emph{either} one dissipation mechanism \emph{or} one
fluid-storage mechanism. For example, the storage $mk$-edge corresponds
to the case of negligible leak-off (impermeable solid, $C'=0$), which
solution is expected to transition from the toughness-dominated $k$-vertex
in the near field to the storage-viscosity dominated $m$-vertex in
the far field. The characteristic lengthscale $\ell_{mk}$ of this
transition can be estimated by evaluating the distance from the tip
where the $k$ and $m$ asymptotes ``intersect'', i.e. $\ell_{k}^{1/2}r^{1/2}\sim\ell_{m}^{1/3}r^{2/3}$
at $r\sim\ell_{mk}$, which gives 
\[
\ell_{mk}=\frac{\ell_{k}^{3}}{\ell_{m}^{2}}=\frac{K'^{6}}{E'^{4}V^{2}\eta'^{2}}
\]

Exploring the other two edges in the similar fashion we get that the
viscosity $m\widetilde{m}$-edge corresponds to negligible solid toughness
($K_{Ic}=0$), and its expected solution transitions from the viscosity-leak-off
dominated $\widetilde{m}$-vertex in the near field to the viscosity-storage
dominated $m$-vertex in the far field. The corresponding transition
lengthscale 
\[
\ell_{m\widetilde{m}}=\frac{\ell_{\widetilde{m}}^{9}}{\ell_{m}^{8}}=\frac{C'^{6}E'^{2}}{V^{5}\eta'^{2}}
\]
Finally, the $\widetilde{m}k$-edge corresponds to the case with negligible
storage ($C'\rightarrow\infty$), and its expected solution transitions
from the toughness-dominated $k$-vertex to the viscosity-leak-off-dominated
$\widetilde{m}$-vertex with the distance from the tip, with the characteristic
transition lengthscale
\[
\ell_{\widetilde{m}k}=\frac{\text{\ensuremath{\ell_{k}^{4}}}}{\ell_{\widetilde{m}}^{3}}=\frac{K'^{8}}{E'^{6}C'^{2}V\eta'^{2}}
\]

It can then be foreseen that for ``small'' leak-off, the solution
is closely approximated by the $mk$-edge
\[
k\underset{\ell_{mk}}{\rightarrow}m\quad(\text{"small" }C')
\]
As the leak-off parameter is increased, the solution trajectory is
dragged towards the leak-off-viscosity $\widetilde{m}$-vertex (while
the near and far fields are invariantly at the $k$ and $m$ vertices,
respectively), such that in the limit of ``large'' leak-off the
solution structure with complete set of the near ($k$), intermediate
$(\widetilde{m}$), and far ($m$) field asymptotes is realized 
\[
k\underset{\ell_{\widetilde{m}k}}{\rightarrow}\widetilde{m}\underset{\ell_{m\widetilde{m}}}{\rightarrow}m\quad(\text{"large" }C')
\]
It is now clear that in the latter ``large'' leak-off case, the
two corresponding transition lengthscales have to separate, i.e.
\[
\ell_{\widetilde{m}k}\ll\ell_{m\widetilde{m}}\quad(\text{"large" }C')
\]
which then suggests a single non-dimensional parameter, $\ell_{m\widetilde{m}}/\ell_{\widetilde{m}k}$,
to quantify the ``smallness'' or ``largeness'' of leak-off, and,
thus, fully parametrize a solution trajectory in the parametric space.
More specifically, we can define non-dimensional leak-off parameter
\[
\chi=\left(\frac{\ell_{m\widetilde{m}}}{\ell_{\widetilde{m}k}}\right)^{1/8}=\frac{C'E'}{V^{1/2}K'}
\]
to parametrize the general HF tip solution (Figure \ref{fig:space}).

\subsubsection{Towards complete HF tip formulation and obtaining the complete tip
solution}

In order to solve the general HF tip problem, which anticipated spatial
fracture and parametric dependence has been addressed above, we need
to appeal to more general description of the crack elasticity than
the Williams' power law class of solutions utilized in so far. That
general elastic framework can be furnished in the form the crack boundary
integral equation, which models the distribution of crack opening
as a ``pile-up'' of dislocations, as discussed in the preceding
chapters.

To briefly recount, we use the solution for a unit gap opened along
semi-infinite line $y=0$, $x>0$ (or $\theta=0$ in polar coordinates),
$u_{\theta}(\theta=0^{+})-u_{\theta}(\theta=0^{-})=1$, (so called
``climb'' dislocation) for the distribution of stress normal to
the dislocation plane $\sigma_{yy}^{D}(x,y=0)=-E'/4\pi x$. Noting
that arbitrary distribution of opening $w(x)$ can be obtained as
pile-up of infinitesimal climb-dislocations $dw(s)$ positioned along
the crack $s\in crack$, the corresponding stress distribution along
the crack plane is obtained by superposition
\begin{figure}[tbh]
\includegraphics[scale=0.25]{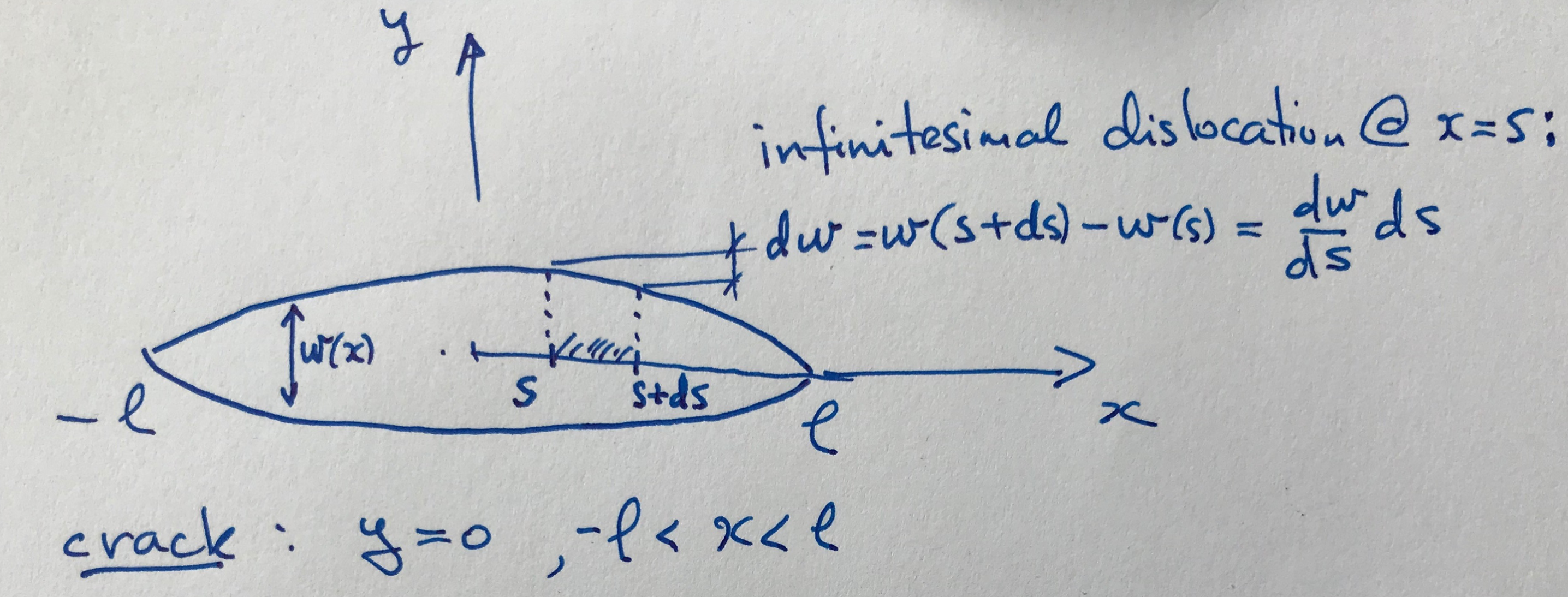}

\caption{Crack as a pile-up of infinitesimal dislocations.\label{fig:pileup}}
\end{figure}

\begin{equation}
\sigma_{yy}(x,y=0)=\int_{crack}\sigma_{yy}^{D}(x-s,y=0)\,\frac{dw}{ds}ds\label{el-1}
\end{equation}

In the particular case of a semi-infinite crack, in the coordinate
frame moving with the crack tip, $r=x_{tip}(t)-x$ for $\theta=\pm\pi$,
the above elasticity equation takes the following form along the crack
\begin{equation}
\sigma_{yy}(r,\theta=\pm\pi)=-p(r)=-\frac{E'}{4\pi}\int_{0}^{\infty}\frac{dw}{ds}\frac{ds}{r-s}\label{elast_tip}
\end{equation}
(this stress expression applies ahead of the crack tip, $\theta=0$,
upon substitution $r\Rightarrow-r$).

General solution for the HF tip is then obtained by solving simultaneously,
the crack elasticity (\ref{elast_tip}), fluid continuity (\ref{balance_leak})
with the Poiseuille law (\ref{lub}), and the propagation condition
(\ref{k}) at $r\rightarrow0$. The numerical solution can be furnished
by a variety of methods, including the method described in \cite{GaDe11}
and, more recently, a method based on Gauss-Chebyshev calculus for
classical (dry) crack problems \cite{ErGu73} applied to HF \cite{ViescaGaragash18}.

\begin{figure}[tbh]
\includegraphics[scale=0.45]{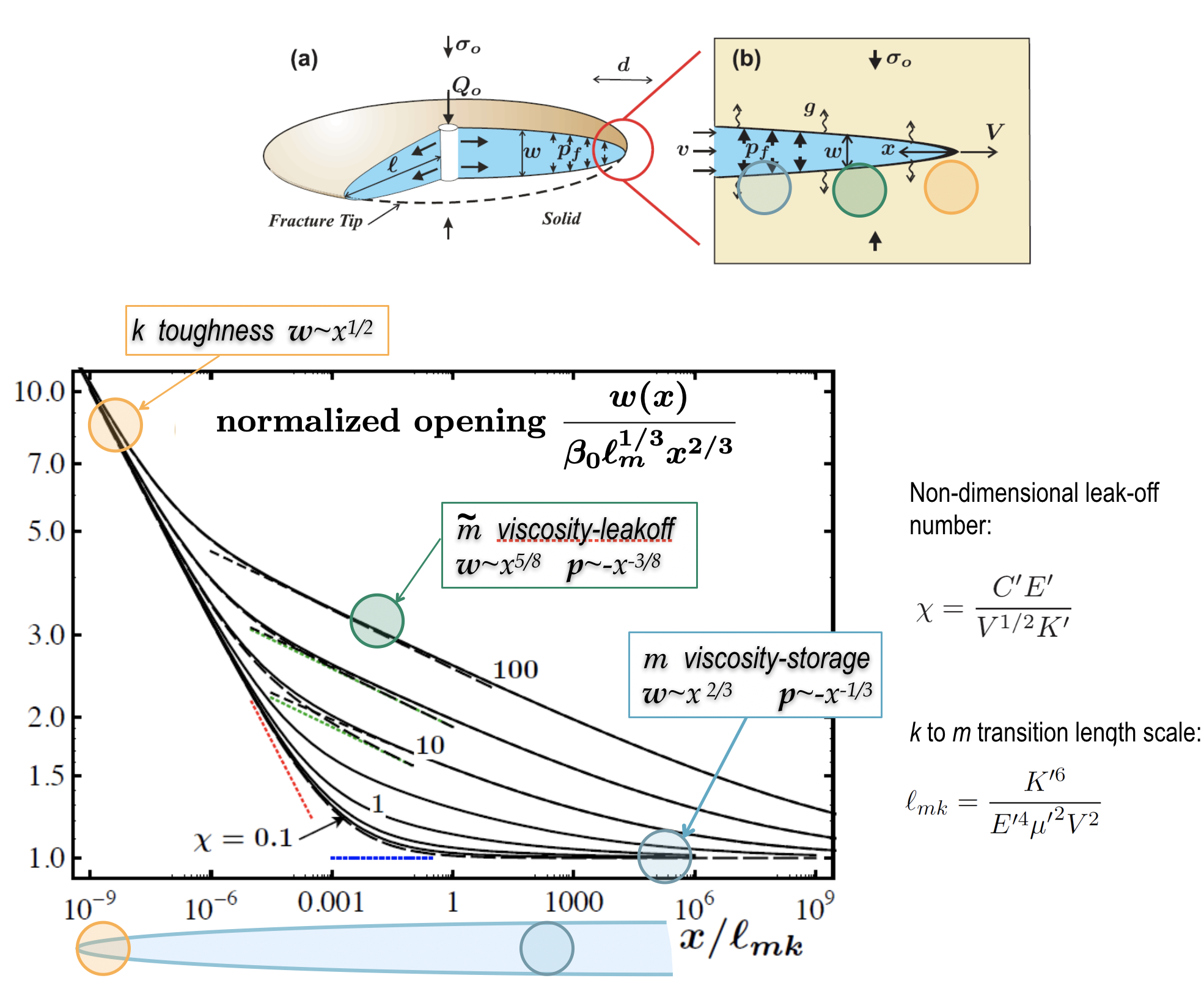}

\caption{Semi-infinite HF general solution for the opening normalized by the
far-field ($m$) asymptote for different values of the non-dimensional
leak-off parameter $\chi$, adopted from \cite{Garagash11}.\label{fig:sol}}
\end{figure}

\subsection{Some physical constraints/limitations on Hydraulic Fracture Tip Model}

The physical limitations of the presented HF tip model are sketched
here, as they emerge from the considerations for the solid and fluid,
respectively. This delineation is somewhat superficial, as the coupling
between the fluid flow and solid rupture is the $O(1)$ effect that
would modify the physical manifestation of the processes.

\subsubsection{Solid}
\begin{itemize}
\item LEFM stress singularity $\Rightarrow$ emergence of a finite non-elastic
region at/ahead the crack tip where the solid undergoes degradation/decohesion.
This fracture ``process zone'' is often modeled as the extension
of the crack along its plane (i.e. material degradation zone is ``collapsed''
for the sake of modeling onto the fracture plane) along which the
cohesive tractions between the nascent fracture surfaces evolve from
the peak material cohesion (at zero fracture opening) to zero cohesion
(at certain critical crack surface separation/opening $w_{c}$).
\item Apparent fracture energy (energy to be spent in creation of a unit
of the \emph{nominal} fracture surface, i.e. projection of the real/rough
surfaces onto the mean plane) is likely not a fixed material property,
but rather scales with the fracture growth
\end{itemize}

\subsubsection{Fluid}
\begin{itemize}
\item Singular fluid suction at the tip of fully-fluid-filled propagating
crack $\Rightarrow$ emergence of a finite fluid lag - a near tip
region filled by fracturing fluid vapor/volatiles (in the case of
impermeable rock) or/and by the pore fluid sucked into the lag from
the surrounding permeable rock \cite{GaragashDetournay00}
\item Coupling of the fluid (lag) and the solid (decohesion) process zones
via the fluid flow \cite{Garagash19}
\item Fracture surface rough topography about nominally-flat fracture plane
may lead to the enhanced dissipation in the viscous fluid flow (departure
from the Pouiseille flow between two parallel, flat plates)
\item Fluid viscous drag on the fracture surfaces in the direction of the
flow (i.e. towards the fracture tip) \cite{WrobelMishuris15}
\end{itemize}

\section{Finite HF of Simple Geometries (homogeneous rock and in-situ stress
field)}

Under conditions of homogeneous (spatially uniform), isotropic rock
properties and homogeneous in-situ stress field, hydraulic fractures
would tend to propagate along a single plane normal to the direction
of the minimum (compressive) in-situ stress, referred to as the fracture
confining stress $\sigma_{o}$. Aside from shallow depth cases, the
minimum in-situ stress is usually horizontal, leading to a vertical
hydraulic fracture plane.

\begin{figure}[tbh]
\includegraphics[scale=0.12]{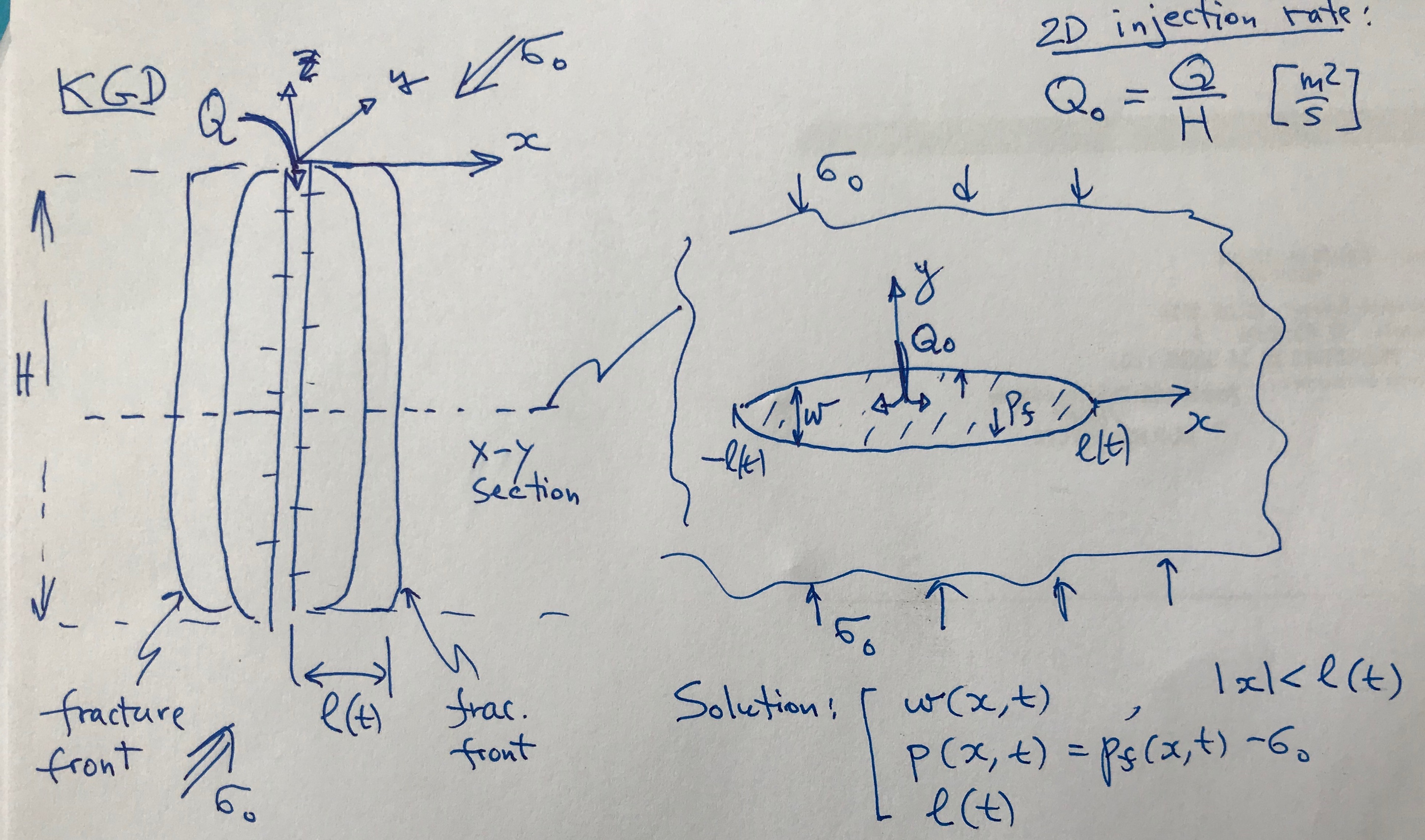}

\caption{KGD fracture model: (left) bi-wing vertical hydraulic fracture of
half-length $\ell(t)$ propagating from the line source with fluid
injection rate $Q$ distributed over height $H\gg\ell$; (tight) horizontal
2D cross-section corresponding to a plane-strain crack driven by 2D
injection rate $Q_{o}=Q/H$.\label{fig:KGD}}
\end{figure}

The expected fracture geometry within its plane is defined by the
dimensionality of the fluid source. In the following, we adopt a fixed
coordinate system $x$ (fracture propagation direction), $y$ (fracture
opening direction), and $z$ (fracture height direction.
\begin{itemize}
\item A circular (\textbf{penny-shape}) fracture geometry as the result
of radial propagation from a point fluid source, approximating a localized
perforated interval of a otherwise cased wellbore. The wellbore orientation
(vertical, horizontal, or inclined) does not reflect on the fracture
geometry in this fluid source idealization. In this case the fracture
dimensions are: the instantaneous fracture radius $\ell(t)$ (distance
from the fluid source to the current location of the circular fracture
front, $r=\ell(t)$ in the vertical x-z fracture plane, in the coordinate
system with the origin at the fluid source) and the opening $w(r,t)$
with $0<r<R(t)$
\item A plane-strain, bi-wing fracture geometry as the result of propagation
from a line fluid source, approximating a spatially extended perforated
interval of a \emph{vertical} wellbore (Figure \ref{fig:KGD}). This
fracture geometry is referred to as \textbf{KGD}-fracture (after Khristianovic,
Zheltov, Geertsma and De Klerk \cite{KhZh55,GeKl69}). The fracture
dimensions are: the instantaneous half-length of the fracture $\ell(t)$
in the (bi-)propagation direction(s) away from the line fluid source
and the opening $w(x,t)$, $-\ell(t)<x<\ell(t)$, with $x=y=0$ giving
the location of the line source. The plane-strain approximation is
the result of taking the third fracture dimension -its height - as
infinite (similarly approximating the line fluid source as infinite),
such that any horizontal cross-section (x-y planes) of the fracture
is identical to another.
\item A third simplified hydraulic fracture geometry commonly considered
is the so-called \textbf{PKN} (after Perkins, Kern and Nordgren \cite{PeKe61,Nord72}),
or fracture with constrained height, which postulates a bi-wing fracture
with a fixed height $h=const$ and half-length $\ell(t)\gg h$. Physically,
the fixed height approximation emerges from the lithological/stress
vertical heterogeneity - where the more competent cap-rock layers,
above and below the hydraulically-stimulated (reservoir) layer of
height $h$, support higher level of confining stress, effectively
suppressing the hydraulic fracture height growth into the cap-rock.
The main modeling approximation emerging from the elongated fracture
geometry ($\ell\gg h$) is the plane-strain approximation for any
vertical cross-section (y-z planes) of the fracture.
\end{itemize}
The above simplified fracture geometry reduce the problem dimensionality
to 1D (in the direction of fracture propagation), allowing for a variety
of (semi) analytical approaches, providing basic understanding of
propagation of finite volume hydraulic fractures and their parametric
regimes, and also serve as benchmarks for numerical methods been developed
for more general treatment of the hydraulic fracture problem (e.g.,
the 2 or 3D geometries, non-uniform and anisotropic properties and
stress, near wellbore effects, etc, see \cite{lecampion2018numerical}
for a recent review).

\subsection{KGD Fracture Formulation}

Field (crack line) equations:
\begin{itemize}
\item Crack elasticity (dislocation pile-up)
\begin{equation}
\sigma_{yy}(x,y=0,t)=-p(x,t)=-\frac{E'}{4\pi}\int_{-\ell(t)}^{+\ell(t)}\frac{\partial w(s,t)}{\partial s}\frac{ds}{x-s}\label{crack_KGD}
\end{equation}
\item Fluid flow (continuity and Poiseuille's law)
\begin{equation}
\frac{\partial w}{\partial t}+\frac{\partial wv_{f}}{\partial x}=g\qquad v_{f}=-\frac{w^{2}}{\eta'}\frac{\partial p}{\partial x}\label{balance_KGD}
\end{equation}
Local rate of fluid exchange with the rock (Carter's leak-off) is
\begin{equation}
g(x,t)=\frac{C'}{\sqrt{t-t_{o}(x)}}\label{g}
\end{equation}
where $t_{o}(x)$ is the arrival time of the crack tip at location
$x$ along the crack plane ($y=0$), i.e.
\begin{equation}
\ell(t_{o}(x))=x\label{arrival}
\end{equation}
\item Boundary condition (solid): crack propagation condition, $K_{I}=K_{Ic}$,
rewritten in terms of the crack tip asymptotics, (\ref{k}),
\begin{equation}
w(x,t)=\frac{K'}{E'}\sqrt{\ell(t)-|x|}\qquad(|x|\rightarrow\ell(t))\label{prop}
\end{equation}
where $K'=\sqrt{32/\pi}K_{Ic}$ toughness parameter is introduced
to absorb the numerical factor.
\item Boundary condition (fluid) at the inlet: constant injection rate $Q_{o}$
{[}m$^{2}$/s{]} imposed at the crack inlet $x=0$, is partitioned
equally between the two fracture wings
\[
\left(wv_{f}\right)_{x=\pm0}=\pm Q_{o}/2
\]
Alternatively, can express the inlet flow boundary condition as the
global fluid volume balance statement for injection volume 
\begin{equation}
V_{inj}=Q_{o}t=V_{crack}+V_{leak}\label{global}
\end{equation}
where the time-dependent crack and cumulative leak-off volumes are
\begin{equation}
V_{crack}=\int_{-\ell(t)}^{+\ell(t)}w(x,t)dx\qquad V_{leak}=\int_{0}^{t}dt'\int_{-\ell(t')}^{+\ell(t')}g(x,t')dx\label{Vcrack}
\end{equation}
The expression for $V_{leak}$ can be further simplified for the Carter's
leak-off case (\ref{g}-\ref{arrival}) by using the crack symmetry,
$\int_{-\ell}^{+\ell}...=2\int_{0}^{\ell}...$, changing the order
of integration $\int_{0}^{t}dt'\int_{0}^{\ell(t')}dx=\int_{0}^{\ell(t)}dx\int_{t_{o}(x)}^{t}dt'$,
and integrating by parts
\begin{equation}
V_{leak}=2C'\int_{0}^{t}\frac{\ell(t')dt'}{\sqrt{t-t'}}\label{Vleak}
\end{equation}
\item Boundary condition (fluid) at the tip(s): fluid velocity at the tip
matches the fracture propagation velocity (fluid neither lags no outpaces
the fracture front)
\begin{equation}
v_{f}(x\rightarrow\pm\ell(t),t)=\pm\frac{d\ell}{dt}\label{fluid_tip}
\end{equation}
\end{itemize}
The above set of equations governs the solution to the KGD hydraulic
fracture propagation, specifically evolution of the net fluid pressure
$p(x,t)$ and crack opening $w(x,t)$ distributions, and the fracture
half-length $\ell(t)$.

As already anticipated in the study of the hydraulic fracture tip
problem, we may expect that the finite hydraulic fracture propagation
(such as KGD) in a general case may fall somewhere within the parametric
space defined by four limiting propagation regimes, dominated by one
of the two dissipation mechanisms (viscosity vs. toughness) and one
of the two fluid storage mechanisms (storage in the fracture vs. porous
rock/leak-off). In the following we will furnish approximate solutions
in these four limiting propagation regimes based on our understanding
of the respective fracture tip asymptotics. We then provide some ideas
about the general KGD solution trajectory with regard to these limiting
regimes. In doing so, we will leave out the details of general scaling
arguments for KGD fracture \cite{Deto04,HuGaragash10} and numerical
solution in general case (which can be readily furnished), but rather
focus on how knowing the limiting regime solutions allows to understand
the general evolution of KGD fracture.

\subsection{Toughness-dominated KGD propagation}

In this case, the fluid viscosity and associated pressure drop in
the fluid flow in the fracture can be neglected, leading to uniform
pressure distribution in the crack
\[
\frac{\partial p}{\partial x}=0\quad\Rightarrow\quad p=p(t)
\]
This reduces the crack elasticity problem to the classical Griffith's
crack framework of LEFM, which solution can be obtained from formal
inversion of the dislocation integral (\ref{crack_KGD}) with uniform
left hand side
\begin{equation}
w(x,t)=\frac{4p(t)}{E'}\sqrt{\ell(t)^{2}-x^{2}}\label{w_tough}
\end{equation}
which together with the tip asymptotics for the propagating crack
(\ref{prop}) leads to the relation between the net pressure and the
crack half-length
\begin{equation}
4p(t)\sqrt{2\ell(t)}=K'\label{p_tough}
\end{equation}
Using (\ref{w_tough}) to evaluate the crack volume in the global
fluid continuity (\ref{global}-\ref{Vleak}), together with and (\ref{p_tough}),
furnishes the evolution equation for crack length
\begin{equation}
Q_{o}t=\underset{V_{crack}}{\underbrace{2\pi\frac{p(t)}{E'}\ell(t)^{2}}}+\underset{V_{leak}}{\underbrace{2C'\int_{0}^{t}\frac{\ell(t')dt'}{\sqrt{t-t'}}}}\quad\text{with}\quad p(t)=\frac{K'}{\sqrt{32\ell(t)}}\label{global_tough}
\end{equation}
The solution of the above non-linear integral equation can be obtained
numerically \cite{Bunger05}, however the two limiting regimes of
this solution corresponding to the negligible leak-off (storage-dominated)
and leak-off dominated cases can be obtained analytically.

\subsubsection{Toughness-storage-dominated regime ($K$-solution)}

In this case, setting $C'=0$ in (\ref{global_tough}) we find
\begin{equation}
\ell(t)=\frac{2}{\pi^{2/3}}\left(\frac{E'Q_{o}t}{K'}\right)^{2/3}\qquad p(t)=\frac{\pi^{1/3}}{8}\left(\frac{K'^{4}}{E'Q_{o}t}\right)^{1/3}\label{K}
\end{equation}
To assess when this storage-dominated regime may be appropriate approximation
of the general solution with non-zero leak-off $C'>0$, let us evaluate
the corresponding leak-off volume
\[
V_{leak@K}\sim C'\left(\frac{E'Q_{o}}{K'}\right)^{2/3}t^{7/6}
\]
(where we have omitted numerical O(1) prefactor for brevity). Comparing
$V_{leak@K}$ to the injected volume $Q_{o}t$, we observe that the
K-solution (toughness-storage-dominated propagation regime) is applicable
at ``early'' times. The notion of ``early'' can be quantified/estimated
by defining transition timescale $t_{*}$ when the solution departs
from the K-solution, i.e. when $V_{leak@K}(t_{*})\sim Q_{o}t_{*}$
\begin{equation}
t_{*}=t_{K\rightarrow\widetilde{K}}=\frac{Q_{o}^{2}K'^{4}}{C'^{6}E'^{4}}\label{t_KK}
\end{equation}
Thus, we anticipate that the storage dominated K-solution (\ref{K})
applies at early times $t\ll t_{*}$.

\subsubsection{Toughness-leak-off regime ($\widetilde{K}$-solution)}

In this case, we neglect the crack volume in the fluid balance equation
(\ref{global_tough}), i.e. write $Q_{o}t\approx V_{leak}(t)$. For
arbitrary time power law for the half-length $\ell(t)\propto t^{\alpha}$,
the leak-off volume is also a power law $V_{leak}\propto t^{\alpha+1/2}$,
suggesting the crack length evolution as the square root of time.
The complete leak-off-dominated $\widetilde{K}$-solution follows
as
\begin{equation}
\ell(t)=\frac{1}{\pi}\frac{Q_{o}}{C'}\sqrt{t}\qquad p(t)=\sqrt{\frac{\pi}{32}}\frac{K'C'^{1/2}}{Q_{o}^{1/2}\,t^{1/4}}\label{K_tilde}
\end{equation}
Similarly to the considerations for the applicability of the storage-dominated
$K$-solution in the above, we can evaluate the (time range of) applicability
of the leak-off dominated solution by comparing the corresponding
crack volume 
\[
V_{crack@\widetilde{K}}\sim\frac{K'}{E'}\left(\frac{Q_{o}}{C'}\right)^{3/2}t^{3/4}
\]
to the injected volume $Q_{o}t$. We observe that the leak-off dominated
$\widetilde{K}$ solution is applicable when $V_{crack@\widetilde{K}}\ll Q_{o}t$,
or at large times $t\gg t_{*}$ with (not surprisingly) the same transition
timescale $t_{*}$, (\ref{t_KK}).

\subsubsection{Summary of the toughness dominated propagation regime}

\begin{figure}[tbh]
\includegraphics[scale=0.07]{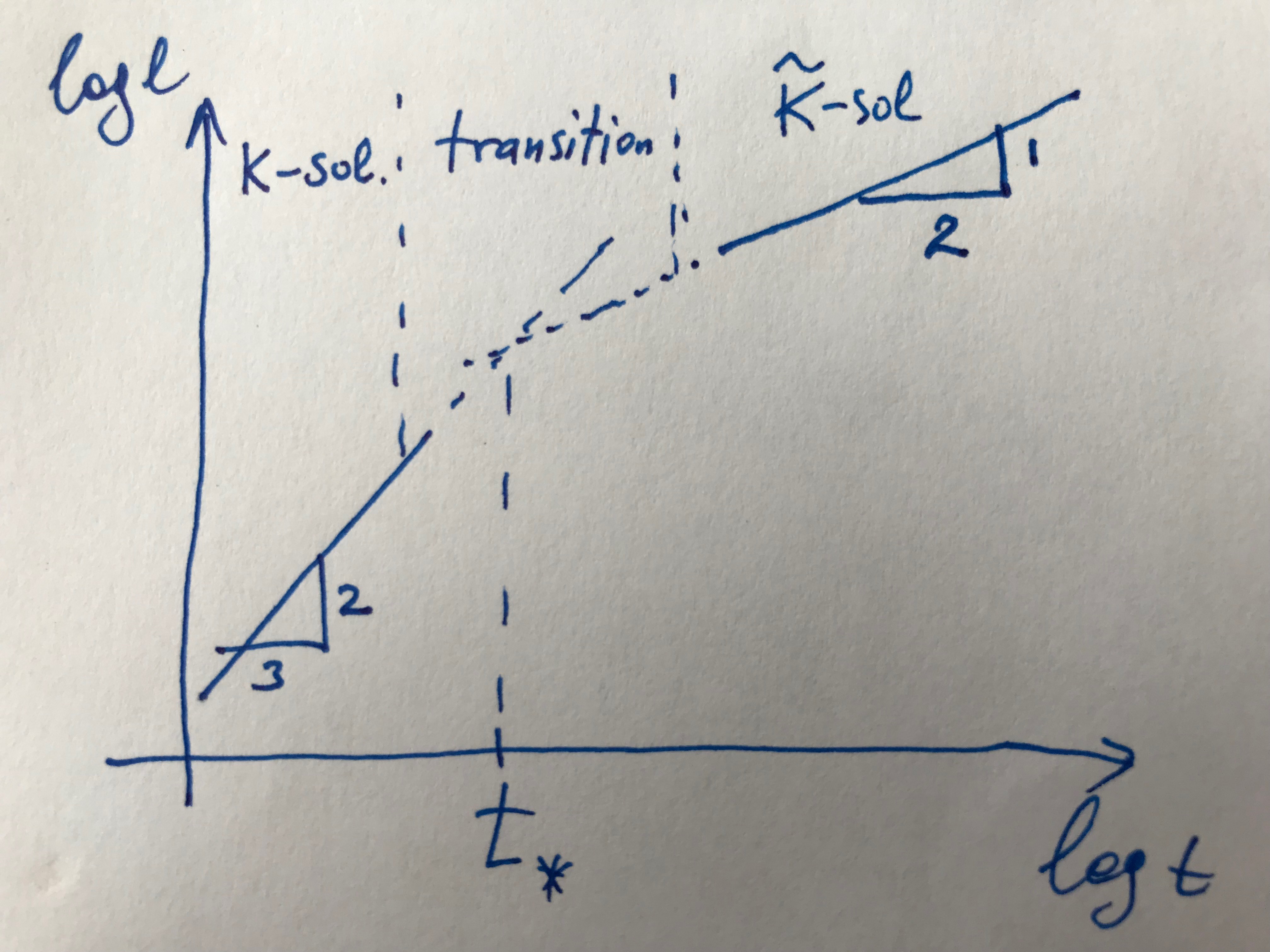}

\caption{KGD fracture propagation in the toughness-dominated regime from early-time
storage-dominated ($K$) to large-time leak-off-dominated ($\widetilde{K}$)
regime.\label{fig:KGD_tough}}
\end{figure}

The solution is given at the early ($t\ll t_{*})$ and large ($t\gg t_{*}$)
time by the storage $K$- and leak-off $\widetilde{K}$- dominated
asymptotes, while the transition from one to the other over $t\sim t_{*}$
can be computed numerically by solving the nonlinear integral equation
(\ref{global_tough}) \cite{Bunger05}, Figure \ref{fig:KGD_tough}.
The transition timescale $t_{*}=t_{K\rightarrow\widetilde{K}}$ is
given by (\ref{t_KK}).

\subsection{Viscosity-dominated KGD propagation}

In this case, the solid toughness is negligible compared to the dissipation
in the viscous fluid flow. (We leave the task of quantifying the corresponding
parametric regime for later). As in the toughness-dominated propagation
in the above, let us focus first on the two sub-regimes, dominated
by the storage in the fracture (negligible leak-off), which will call
$M$-solution, and by the leak-off (negligible fluid storage in the
fracture), which will call $\widetilde{M}$-solution, respectively.
Intuitively, we expect that the former and the latter would actually
define the early and large time behavior of the viscosity-dominated
propagation.

\subsubsection{Viscosity-storage-dominated ($M$) solution}

As we have established from the fracture tip solution, the corresponding
tip behavior is given by the storage--viscosity-dominated asymptote
(\ref{m}), $w\propto r^{2/3}$ and $p\propto-r^{-1/3}$ where the
distance from the (right) tip $r=\ell(t)-x$. One can use a numerical
method to find the complete solution for the KGD fracture in this
regime, e.g. \cite{AdDe02}. Instead, we will explore a simple analytical
approximation of the $M$-solution, ``inspired'' by the $m$ tip
asymptote, alone the lines of a more elaborate approach taken in \cite{Dontsov16,Dontsov17}.
Specifically, we extend the opening tip asymptote to provide a consistent
approximation for the entire KGD crack via replacing $r\Rightarrow(\ell^{2}-x^{2})/2\ell$,
which (i) retains the tip asymptotic form in the limit $x\rightarrow\pm\ell$,
and (b) allows for KGD crack symmetry and closure at both tips
\begin{equation}
w(x,t)\approx\delta_{m}\varphi(2/3)\,\ell_{m}^{1/3}\left(\frac{\ell^{2}-x^{2}}{2\ell}\right)^{2/3}\label{wM}
\end{equation}
where $\delta_{m}\varphi(2/3)=2^{1/3}3^{5/6}$ is the numerical prefactor
and the tip lengthscale $\ell_{m}$ is evaluated in terms of the instantaneous
crack tip velocity $V=d\ell/dt$
\begin{equation}
\ell_{m}=\frac{\eta'}{E'}\frac{d\ell}{dt}\label{lmM}
\end{equation}
Corresponding approximation for the net-pressure can be evaluated
from the crack elasticity integral (\ref{crack_KGD}). To complete
the solution, we use the global fluid balance, which in the storage-dominated
case with approximate opening given by (\ref{wM}) reduces to 
\[
Q_{o}t=V_{crack}=\int_{-\ell}^{+\ell}wdx\approx\frac{2^{1/3}3^{5/6}}{2^{2/3}}\,\ell_{m}^{1/3}\ell^{5/3}\int_{-1}^{+1}\left(1-\xi^{2}\right)^{2/3}d\xi=\alpha\,\ell_{m}^{1/3}\ell^{5/3}\quad\text{with}\quad\alpha\approx2.93
\]
(where $x/\ell$ was denoted as $\xi$ in the above opening integral).
In view of (\ref{lmM}), the above equation provides a simple ODE
to solve for the crack half-length
\begin{equation}
\ell(t)\approx\underset{\gamma_{m}\approx0.625}{\underbrace{\left(\frac{3}{2\alpha^{3}}\right)^{1/6}}}\underset{L_{m}(t)=\text{dynamic lengthscale}}{\underbrace{\left(\frac{E'Q_{o}^{3}}{\eta'}\right)^{1/6}t^{2/3}}}\label{lM}
\end{equation}

This approximate solution satisfies all governing equations of the
KGD crack with exception for the local (point-wise) fluid flow equation
(only the global fluid balance, and not the local form, has been used
in the construction of the solution). Notwithstanding, comparison
with the accurate numerical solution of complete set of KGD equations
for the non-dimensional crack length $\gamma_{m}=0.6152$ \cite{AdDe02},
shows a good agreement (1\% error).

As already a familiar exercise, we can assess when does the storage-viscosity-dominated
$M$-solution hold when the leak-off is non-negligible by evaluating
$V_{leak@M}$ and comparing it with the injected volume. This predictably
yields that the storage-dominated solution corresponds to the early
time solution, when $t\ll t_{*}$, with the transition timescale given
by ($V_{leak@M}(t_{*})\sim Q_{o}t_{*}$)

\begin{equation}
t_{*}=t_{M\rightarrow\widetilde{M}}=\frac{Q_{o}^{3}\eta'}{C'^{6}E'}\label{t_MM}
\end{equation}

\subsubsection{Viscosity-leak-off-dominated ($\widetilde{M}$) solution}

As we have established from the fracture tip solution, the corresponding
tip behavior is given by the leak-off--viscosity-dominated asymptote
(\ref{m_tilde}), $w\propto r^{5/8}$ and $p\propto-r^{-3/8}$ where
the distance from the (right) tip $r=\ell(t)-x$. Following the same
route to the approximate $\widetilde{M}$-solution inspired by the
tip asymptote, we write for the crack opening 
\begin{equation}
w(x,t)\approx\delta_{\widetilde{m}}\varphi(5/8)\,\ell_{\widetilde{m}}^{3/8}\left(\frac{\ell^{2}-x^{2}}{2\ell}\right)^{5/8}\label{wM_tilde}
\end{equation}
where $\delta_{\widetilde{m}}\varphi(5/8)$ is the numerical prefactor
and the tip lengthscale $\ell_{\widetilde{m}}$ is evaluated in terms
of the instantaneous crack tip velocity $V=d\ell/dt$
\begin{equation}
\ell_{\widetilde{m}}=\left(\frac{C'\eta'}{E'}\right)^{2/3}\left(\frac{d\ell}{dt}\right)^{1/3}\label{lmM_tilde}
\end{equation}
Corresponding approximation for the net-pressure can be evaluated
from the crack elasticity integral (\ref{crack_KGD}). To complete
the solution, we use the global fluid balance, which in the leak-off-dominated
case reduces to $Q_{o}t=V_{leak}$, which has been already solved
for the crack -length evolution in the toughness-leak-off-dominated
case in the above, with the solution given by 
\begin{equation}
\ell(t)=\frac{1}{\pi}\frac{Q_{o}}{C'}\sqrt{t}\label{lM_tilde}
\end{equation}
This, in fact, suggests that the fracture evolution in the leak-off
dominated case is irrespective of the solid vs. viscous fluid dissipation
partition and given universally by the above expression. Substituting
the crack length expression into $\ell_{\widetilde{m}}$ and the result
into (\ref{wM_tilde}) completes the $\widetilde{M}$-solution.

Evaluating $V_{crack@\widetilde{M}}$ and comparing it with the injected
volume renders the viscosity-leak-off-dominated $\widetilde{M}$-solution
as a large time asymptote valid when $t\gg t_{*}$ with the transitional
timescale defined by (\ref{t_MM}).

\subsubsection{Summary of the viscosity-dominated propagation regime}

The solution is given at the early ($t\ll t_{*})$ and large ($t\gg t_{*}$)
time by the storage $M$- and leak-off $\widetilde{M}$- dominated
asymptotes, while the transition from one to the other over $t\sim t_{*}$
can be computed numerically by solving the complete set of the KGD
governing equations with $K'=0$ \cite{AdDe08,HuGaragash10}. The
transition timescale $t_{*}=t_{M\rightarrow\widetilde{M}}$ is given
by (\ref{t_MM}).

\subsection{General KGD solution, Parametric Space, and Perspective from the
Tip}

\begin{figure}[tbh]
\includegraphics[scale=0.4]{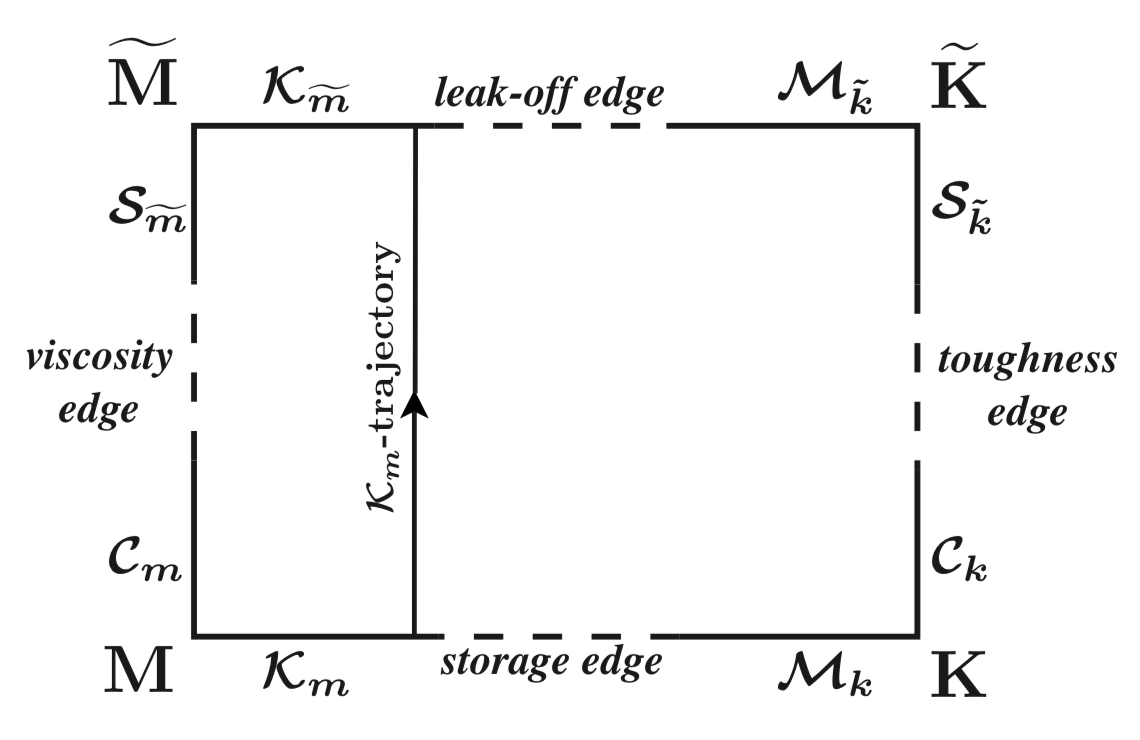}

\caption{KGD fracture parametric space and a solution trajectory parametrized
by, e.g., toughness number $K_{M}$ and leak-off number $C_{M}$,
adopted from \cite{HuGaragash10}.\label{fig:KGD_rectagle}}
\end{figure}

General KGD solution can be envisioned within the rectangular parametric
space $M\widetilde{M}\widetilde{K}K$, Figure \ref{fig:KGD_rectagle},
with vertices corresponding to the above four limiting regime solutions,
while the edges corresponding to the viscosity-dominated $M\widetilde{M}$
($K'=0$), toughness-dominated $K\widetilde{K}$ ($\eta'$=0), storage-dominated
$MK$ ($C'=0$), and leak-off-dominated $\widetilde{M}\widetilde{K}$
($C'\rightarrow\infty$) regimes. \cite{Deto04,HuGaragash10}

So far we have sketched-out the solution along the $M\widetilde{M}$
and the $K\widetilde{K}$ edges, corresponding to the fracture evolving
from the storage to the leak-off dominated state over the transient
timescale $t_{*}$ (given by $t_{M\rightarrow\widetilde{M}}$ and
$t_{K\rightarrow\widetilde{K}}$, respectively). Thus, the solution
trajectory along these two edges can be parametrized by corresponding
normalized time $t/t_{*}$, which can alternatively be rewritten in
terms of the the non-dimensional leak-off numbers
\[
C_{M}=\left(\frac{t}{t_{M\rightarrow\widetilde{M}}}\right)^{1/6}=C'\left(\frac{E'\,t}{Q_{o}^{3}\eta'}\right)^{1/6}\qquad C_{K}=\left(\frac{t}{t_{K\rightarrow\widetilde{K}}}\right)^{1/6}=C'\left(\frac{E'^{4}t}{Q_{o}^{2}K'^{4}}\right)^{1/6}
\]
Put in other words, the departure of the viscosity- (toughness-) dominated
solution from the corresponding storage-dominated early time vertex
$M$ ($K$) can be parametrized by the time-dependent leak-off number
$C_{M}$ ($C_{K}$) - the quantifiable measure of the relative importance
of the fracturing fluid leak-off vs. fluid storage in the KGD fracture.

Here we can relate this evolution of the KGD fracture to the evolution
of its near tip field. Recall, for the viscosity-dominated case ($K'=0$),
that the HF tip solution corresponds to the transition with distance
from the tip $r$ from the leak-off dominated near field $\widetilde{m}$-solution
to the storage-dominated far field $m$-solution over the characteristic
distance $r\sim\ell_{m\widetilde{m}}=C'^{6}E'^{2}/V^{5}\eta'^{2}$.
Intuitively, the finite fracture (KGD) fracture solution should be
storage-dominated if that near tip transition can be accomplished
within the scale of the finite fracture of half-length $\ell$, i.e.
when $\ell_{m\widetilde{m}}\ll\ell$, and, leak-off dominated otherwise,
when $\ell_{m\widetilde{m}}\gg\ell$. Evaluating the tip-to-global
fracture lengthscale ratio $\ell_{m\widetilde{m}}/\ell$ assuming
the storage-dominated KGD solution (\ref{lM}) for $\ell(t)$ and
the corresponding crack tip velocity $V(t)=d\ell/dt$, we find (while
dropping numerical prefactors of $O(1)$)
\[
\left(\frac{\ell_{m\widetilde{m}}}{\ell}\right)_{@M}\sim C_{M}^{6}(t)
\]
Since the non-dimensional leak-off number $C_{M}$ is increasing power
law of time, we confirm the succession of the propagation regime in
time from the storage to the leak-off dominated as seen from the perspective
of the evolving fracture tip behavior corresponding to the dynamically
expanding near tip transition lengthscale until it is no longer small
compared to the finite fracture half-length $\ell$.

To glean the behavior along the energy dissipation axis of the parametric
space (e.g. along the storage $MK$ edge of the parametric space),
let us look at the storage-dominated tip solution corresponding to
the transition from the toughness dominated near field $k$-solution
to the viscosity-dominated far-field $m$-solution over the characteristic
distance from the tip $r\sim\ell_{mk}=K'^{6}/E'^{4}V^{2}\eta'^{2}$.
Once again, the expectation is that the finite fracture solution is
dominated by viscosity (viscous dissipation) if the near tip ($k$
to $m$) transition can be accomplished within the scale of the finite
fracture half-length, i.e. when $\ell_{mk}\ll\ell$, and toughness-dominated
otherwise, when $\ell_{mk}\gg\ell$. Evaluating the corresponding
lengthscale ratio $\ell_{mk}/\ell$ assuming the viscosity-dominated
($M$) KGD solution (\ref{lM}) for $\ell(t)$ and $V(t)$, we find
\[
\left(\frac{\ell_{mk}}{\ell}\right)_{@M}\sim K_{M}^{6}\qquad\text{with}\qquad K_{M}=\frac{K'}{E'^{3/4}Q_{o}^{1/4}\eta'^{1/4}}
\]
where $K_{M}$ has the meaning of the non-dimensional toughness reflecting
the partition of energy dissipation between the solid and fluid. Similar
exercise along the leak-off edge $\widetilde{M}\widetilde{K}$ of
the parametric space results in comparing the near tip ($k$ to $\widetilde{m}$)
transition lengthscale to the fracture half-length in the context
of the viscosity-dominated ($\widetilde{M}$) KGD solution yields
\[
\left(\frac{\ell_{\widetilde{m}k}}{\ell}\right)_{@\widetilde{M}}\sim K_{\widetilde{M}}^{6}\qquad\text{with}\qquad K_{\widetilde{M}}=K_{M}
\]

Since $K_{M}$ (and $K_{\widetilde{M}}$) is time-independent, we
conclude that the dissipation partition in the KGD fracture is time-invariant,
and specifies a $K_{M}=const$ trajectory for a general KGD solution
in the paramedic space, corresponding to the evolution of the propagating
fracture from the early time storage to the large time leak-off dominated
regime at a fixed dissipation partition.

\subsection{Concluding remarks}

We have focused on exploring the solution to the KGD (plane strain)
hydraulic fracture and its parametric dependence in some details in
the above. We showed that
\begin{itemize}
\item Approximate solutions in the limiting propagation regimes (when one
fluid storage mechanism \emph{and} one dissipation process dominate)
can be effectively constructed from the corresponding near tip limiting
solutions ``adapted'' to the finite fracture geometry and satisfying
the global fluid balance equation
\item In doing so, it is apparent that the near tip fracture behavior exerts
the 1st order control on the overall solution of a finite HF
\item The general finite HF solution can be represented as a trajectory
in parametric space $M\widetilde{M}\widetilde{K}K$, defined by the
four vertices corresponding to the above limiting regime solutions.
The fracture evolves with time from the storage dominated to the leak-off
dominated regime (tracked by the non-dimensional leak-off parameter
$C_{M}\propto t^{1/6}$), while, for the KGD fracture, the dissipation
partition between the solid (toughness) and the fluid (viscous friction
losses in the fluid flow) dissipation is invariant (quantified by
the time-independent non-dimensionless toughness parameter $K_{M}$)
\item The non-dimensional leak-off $C_{M}$ and toughness $K_{M}$ parameters
which instantaneous values define the current fracture propagation
regime appear naturally when the dynamic HF tip transition lengthscale
(e.g., in case of $C_{M}$,- $\ell_{m\widetilde{m}}$ for the transition
with distance from the tip from the leak-off to storage dominated
viscous regime) is compared to the lengthscale (half-length $\ell)$
of the finite fracture. So, once again the evolving structure of the
near tip solution with changing fracture propagation velocity largely
governs the evolution of the finite hydraulic fracture.
\end{itemize}
The radial (penny-shape) hydraulic fracture propagation from a point-fluid
source can be treated in the similar manner to that of the KGD, once
the crack elasticity and fluid flow equations are appropriately updated
to the axisymmetric geometry. A general solution for the penny-shape
HF can furnished in the same parametric space $M\widetilde{M}\widetilde{K}K$,
with notable difference that, in this case, both the non-dimensional
leak-off $C_{M}$ and the non-dimensional toughness $K_{M}$ are increasing
power laws of time (e.g., \cite{Detournay16}). This suggest that
a general solution trajectory would start from the storage-viscosity
and will end up in the leak-off-toughness dominated regime, as opposed
to the KGD fracture propagation which preserves the dissipation partitioning.

\bibliographystyle{unsrt}
\bibliography{bibrefs_garagash_Aug21.bib}

\end{document}